\documentclass[aps,prb,reprint,superscriptaddress,nofootinbib]{revtex4-2}

\usepackage{graphicx}
\usepackage{amsmath,amssymb,bm}
\usepackage{xcolor}
\usepackage{booktabs}

\usepackage[
colorlinks=true,
linkcolor=blue,
citecolor=blue,
urlcolor=blue
]{hyperref}

\newcommand{\dx}{d_{x^2-y^2}}
\newcommand{\dz}{d_{3z^2-r^2}}

\newcommand{\ii}{\mathrm{i}}

\begin{document}

\title{Interplay of electric field and electronic correlations in pairing competition in bilayer nickelates}
\author{Yanmei Cai}
\affiliation{School of Physics and Astronomy, Beijing Normal University, and Key Laboratory of Multiscale Spin Physics (Beijing Normal University), Ministry of Education, Beijing 100875, China\\}
\author{Yicheng Xiong}
\affiliation{School of Physics and Astronomy, Beijing Normal University, and Key Laboratory of Multiscale Spin Physics (Beijing Normal University), Ministry of Education, Beijing 100875, China\\}
\author{Ying Liang}
\affiliation{College of Physics, Hebei Normal University, and Hebei Advanced Thin Films Laboratory, Shijiazhuang 050024, China\\}
\affiliation{School of Physics and Astronomy, Beijing Normal University, and Key Laboratory of Multiscale Spin Physics (Beijing Normal University), Ministry of Education, Beijing 100875, China\\}
\author{Tianxing Ma}
\email{txma@bnu.edu.cn}
\affiliation{School of Physics and Astronomy, Beijing Normal University, and Key Laboratory of Multiscale Spin Physics (Beijing Normal University), Ministry of Education, Beijing 100875, China\\}

\date{\today}

\begin{abstract}
Layer polarization provides a means of tuning the competition between interlayer and intralayer pairing in bilayer nickelates. Using ground-state constrained-path quantum Monte Carlo, we study interlayer $s^{\pm}$ and intralayer $d$ pairing in a bilayer two-orbital Hubbard model with a layer potential-energy difference $\varepsilon$ representing an effective perpendicular electric field. Within the parameter range studied, increasing $\varepsilon$ generally enhances the finite-size pairing vertex contribution $V_d$ and reduces $V_{s^{\pm}}$, shifting the relative pairing preference toward the $d$-wave channel. Increasing the onsite repulsion enhances both vertex contributions, more strongly in the $d$-wave channel, whereas increasing the crystal-field splitting reduces both and makes $s^{\pm}$ pairing relatively more competitive. Hund's coupling increases the relative competitiveness of $s^{\pm}$ pairing at weak bias, while enhancing the relative strength of $d$-wave pairing at large bias. The bias primarily redistributes charge between layers within each orbital sector, accompanied by an increase in the antiferromagnetic spin structure factor. Hund's coupling also enhances local interorbital parallel-spin correlations. These observations suggest possible connections between charge redistribution, magnetic correlations, and pairing competition, offering insight into how layer polarization and multiorbital interactions influence the relative pairing tendencies in the bilayer nickelate model.
\end{abstract}

\maketitle

\section{Introduction}
In recent years, nickelate superconductors have become an important platform for exploring unconventional superconductivity alongside cuprates and iron-based superconductors. In particular, the bilayer Ruddlesden--Popper nickelate La$_3$Ni$_2$O$_7$ exhibits superconducting signatures with a transition temperature approaching 80~K under high pressure~\cite{Sun2023,ZhangNatPhys2024,JiangPressure2025}, stimulating extensive experimental and theoretical investigations of its electronic structure and pairing mechanism. Structural studies further show that pressure strongly modifies the bilayer geometry and the electronic states associated with the Ni $e_g$ orbitals~\cite{GeislerStruct2024}. Subsequent studies have reported superconducting signatures in ambient-pressure La$_3$Ni$_2$O$_7$-based thin films through strain engineering and optimized growth conditions, with onset temperatures further enhanced by improving film quality and oxygenation control~\cite{Ko2025,Zhou2025,Zhou2026,Osada2025}. In addition, carrier and oxygen tuning has revealed a superconducting dome and associated changes in normal-state transport properties~\cite{WangDome2026}. Spectroscopic and optical studies have shown that the low-energy electronic structure is strongly orbital dependent and sensitive to correlations, strain, and carrier distribution~\cite{YangARPES2024,ChenRIXS2024,WangPRX2026,LiuOptics2024}. These experimental developments indicate that bilayer nickelates provide a tunable multiorbital system in which structural, orbital, and electronic degrees of freedom are closely connected.

The superconducting pairing symmetry and microscopic pairing mechanism of La$_3$Ni$_2$O$_7$ remain under discussion. Its low-energy electronic structure involves both Ni $\dx$ and $\dz$ orbitals and a sizable interlayer hopping in the $\dz$ sector, while correlation studies also emphasize strong orbital selectivity and multiorbital effects~\cite{Luo2023,Christiansson2023,Lechermann2023,Shilenko2023}. Several calculations have proposed sign-changing $s^{\pm}$ pairing associated with interlayer or $\dz$-dominated magnetic processes~\cite{Yang2023,LiuYB2023,QinYang2023,Sakakibara2024,Qu2024,LuoTJ2024,ZhangTheory2024,Tian2024}, while tensor-network calculations of effective bilayer models have also found robust interlayer pairing correlations~\cite{Shen2023,ChenTensor2024}. Other approaches find $d$-wave states that either compete closely with $s^{\pm}$ pairing or emerge as the leading instability. Calculations combining the local density approximation with the fluctuation-exchange approximation found competing $d_{xy}$ and $s^{\pm}$ pairing instabilities~\cite{Heier2024}, while other strong-coupling and many-body studies have proposed distinct pairing scenarios~\cite{Jiang2024,Ryee2024,ZhangBerry2024}. Cluster quantum Monte Carlo calculations emphasized interlayer pairing associated with $\dz$-dominated interlayer spin fluctuations~\cite{Maier2026}, and recent random phase approximation calculations reported that the balance between $d$-wave and sign-changing $s$-wave states depends on interaction parameters and the underlying electronic structure~\cite{XuGuterding2025}. First-principles-based weak-coupling calculations further showed that a modest change in the Ni-$e_g$ crystal-field splitting can reorder competing pairing eigenmodes~\cite{Xia2025}. Constrained-path quantum Monte Carlo (CPMC) calculations in the absence of an interlayer potential bias have also found that Hund's coupling and crystal-field splitting can drive a crossover from dominant intralayer $d$-wave to interlayer $s^{\pm}$-wave pairing correlations~\cite{Xiong2026}. Together, these studies highlight the sensitivity of pairing competition to the orbital electronic structure and multiorbital interactions, motivating the exploration of external control over the relative strengths of competing pairing channels.

A perpendicular electric field offers one possible method of such control by inducing a potential difference between the two layers, thereby modifying layer polarization and orbital occupations. Previous studies using strong-coupling mean-field theory, the density matrix renormalization group, and dynamical cluster approaches have suggested that this layer potential difference can modify the competition between $s^{\pm}$- and $d$-wave pairing channels~\cite{Shao2026,Wei2026,Jia2026}. 

Building on these studies, we investigate how a finite interlayer potential bias modifies the pairing competition and how this response depends on crystal-field splitting, onsite repulsion, and Hund's coupling. We employ the CPMC method to examine these effects in the ground state, treating electronic correlations and quantum fluctuations beyond static mean-field approximations through stochastic imaginary-time projection. To examine the electronic changes accompanying the competition between interlayer $s^{\pm}$-wave and intralayer $d$-wave pairing, we analyze orbital occupations, magnetic correlations, Hund-related interorbital correlations, and the noninteracting susceptibility alongside the pairing correlations. These quantities allow us to explore possible connections among the observed pairing trends, charge redistribution, and spin correlations. Our results characterize how an interlayer potential bias and multiorbital interactions affect the relative strengths of the two pairing channels within the bilayer model studied here and offer a theoretical reference for future experimental studies of electric-field and interface-based control of nickelate superconductivity.

\section{Model and methods}

We consider a bilayer square lattice with two orbitals per Ni site,  denoted by $x\equiv d_{x^2-y^2}$ and $z\equiv d_{3z^2-r^2}$. The Hamiltonian is
\begin{equation}
	\begin{aligned}
		H={}&-\sum_{\langle ij\rangle,\ell,\sigma,\alpha}
		t_1^\alpha
		\bigl(c^\dagger_{i\ell\alpha\sigma}
		c_{j\ell\alpha\sigma}+\mathrm{H.c.}\bigr)
		\\
		&-t_{\mathrm{hyb}}\sum_{i,\ell,\sigma}
		\Bigl[c^\dagger_{i\ell x\sigma}
		\bigl(c_{i+\hat{x},\ell z\sigma}
		-c_{i+\hat{y},\ell z\sigma}\bigr)
		+\mathrm{H.c.}\Bigr]
		\\
		&-t_\perp^z\sum_{i,\sigma}
		\bigl(c^\dagger_{i1z\sigma}c_{i2z\sigma}
		+\mathrm{H.c.}\bigr)
		-\mu\sum_{i,\ell,\alpha,\sigma}n_{i\ell\alpha\sigma}
		\\
		&+U\sum_{i,\ell,\alpha}
		n_{i\ell\alpha\uparrow}n_{i\ell\alpha\downarrow}
		+\frac{\Delta E}{2}\sum_{i,\ell,\sigma}
		\bigl(n_{i\ell x\sigma}-n_{i\ell z\sigma}\bigr)
		\\
		&+\sum_{i,\ell,\sigma,\sigma'}
		\bigl(U'-J_H\delta_{\sigma\sigma'}\bigr)
		n_{i\ell x\sigma}n_{i\ell z\sigma'}
		\\
		&+\frac{\varepsilon}{2}\sum_{i,\alpha,\sigma}
		\bigl(n_{i1\alpha\sigma}-n_{i2\alpha\sigma}\bigr).
	\end{aligned}
	\label{eq:H}
\end{equation}

\begin{figure}[t]
	\centering
	\includegraphics[width=0.48\textwidth]{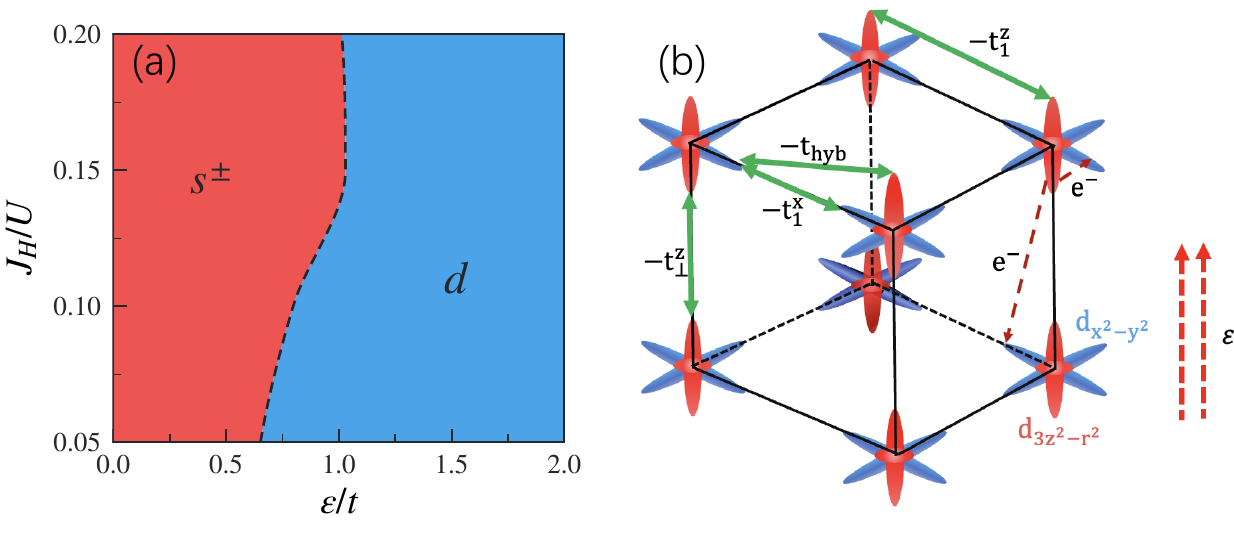}
	\caption{(a) Phase diagram of the dominant pairing tendency in the $(J_H/U,\varepsilon)$ parameter plane at $U/t = 3$,  $\Delta E/t=0.8$, and  $L=8$ and (b) schematic illustration of the bilayer two-orbital model. The dominant hopping processes include the intralayer $\dx$ and $\dz$ hoppings, the sign-changing in-plane interorbital hybridization $t_{\rm hyb}$, and the interlayer $\dz$ hopping $t_\perp^z$.  The layer potential difference $\varepsilon$ mimics a perpendicular electric field.}
	\label{fig:model}
\end{figure}

Here $\ell=1,2$ labels the layers. We impose the relation $U'=U-2J_H$ and retain the density-density form of the Hund interaction. The parameter $\varepsilon$ represents an effective potential difference between the two layers of a single bilayer, mimicking a perpendicular electric field. The physical processes represented by the different parameters in the model are illustrated in Fig.~\ref{fig:model}(b).                                                                                                                                                                                                                                            

The direct interlayer hopping between the $d_{x^2-y^2}$ orbitals, $t_{\perp}^{x}$, is much smaller than the other relevant hopping amplitudes and is neglected in the present model. The dimensionless hopping parameters adopted here are based on the bilayer two-orbital low-energy model obtained from first-principles calculations combined with Wannier downfolding~\cite{Luo2023}. We take $t\equiv t_1^x$ as the energy unit and use
\begin{equation}
\frac{t_1^z}{t}=0.228,\qquad
\frac{t_\perp^z}{t}=1.315,\qquad
\frac{t_{\rm hyb}}{t}=-0.495.
\end{equation}
The average filling is fixed at $\langle n\rangle=0.75$ electrons per orbital, summed over both spin species, corresponding to an average of $1.5$ $e_g$ electrons per Ni site. Unless otherwise stated, the calculations are performed on an $L=8$ bilayer lattice with periodic boundary conditions.

To characterize the competition between intralayer and interlayer singlet pairing, we define the nearest-neighbor $d_{x^2-y^2}$ pair creation operator in the $x$ orbital,
\begin{align}
\Delta_d^{\dagger}(i,\ell)=\sum_{\delta} f_d(\delta)
\big(&c^{\dagger}_{i\ell x\uparrow}c^{\dagger}_{i+\delta,\ell x\downarrow}
-c^{\dagger}_{i\ell x\downarrow}c^{\dagger}_{i+\delta,\ell x\uparrow}\big),
\end{align}
where $f_d(\pm\hat{x})=+1$ and $f_d(\pm\hat{y})=-1$. The interlayer singlet operator is
\begin{equation}
\Delta_{s^{\pm}}^{\dagger}(i)= c^{\dagger}_{i1z\uparrow}c^{\dagger}_{i2z\downarrow} -c^{\dagger}_{i1z\downarrow}c^{\dagger}_{i2z\uparrow}.
\end{equation}
In the bonding/antibonding basis this vertical singlet corresponds to opposite signs between the two bilayer combinations and is therefore denoted $s^{\pm}$.

For a pairing channel $\Gamma$, the equal-time correlation function is
\begin{equation}
	C_{\Gamma}(\bm R)
	=\frac{1}{N_s}\sum_{i,\ell}
	\left\langle
	\Delta_{\Gamma}^{\dagger}(i+\bm R,\ell)
	\Delta_{\Gamma}(i,\ell)
	\right\rangle.
\end{equation}

We subtract the disconnected single-particle contribution and define the interaction-induced vertex contribution
\begin{equation}
V_{\Gamma}(\bm R)=C_{\Gamma}(\bm R)-C_{\Gamma}^{(0)}(\bm R),
\qquad
V_{\Gamma}=\sum_{\bm R}V_{\Gamma}(\bm R).
\end{equation}
A larger $V_{\Gamma}$ indicates a stronger effective pairing tendency in channel $\Gamma$. Because the calculations are performed on a finite lattice, $V_{\Gamma}$ is not identified with a thermodynamic superconducting order parameter. To compare the leading tendencies we use
\begin{equation}
\Delta V=V_d-V_{s^{\pm}},
\end{equation}
so that $\Delta V>0$ ($<0$) indicates a stronger $d$-wave ($s^{\pm}$) tendency.

To further examine the microscopic origin of the pairing competition under an electric-field potential difference, we also evaluate charge distributions and magnetic correlations resolved by layer and orbital. The density for each layer and orbital is defined as
\begin{equation}
n_{\ell\alpha}=\frac{1}{L^2}\sum_{i,\sigma}\langle n_{i\ell\alpha\sigma}\rangle,
\end{equation}
where $\ell=1,2$ and $\alpha=x,z$. This quantity is used below to track field-induced charge redistribution between the two layers and the two orbital sectors. The equal-time spin structure factor is
\begin{equation}
	S^{zz}(\bm q)=\frac{1}{N_s}\sum_{ij}e^{\ii\bm q\cdot(\bm R_i-\bm R_j)}\langle S_i^zS_j^z\rangle,
\end{equation}
where $i,j$ label in-plane positions, $S_i^z=\sum_{\ell,\alpha}(n_{i\ell\alpha\uparrow}-n_{i\ell\alpha\downarrow})$,
and $N_s=2L_xL_y$ is the total number of Ni sites in the bilayer. This definition selects the layer-even ($q_z=0$) channel, with particular emphasis on the in-plane antiferromagnetic wave vector $\bm M=(\pi,\pi)$. To characterize the local spin alignment between the two $e_g$ orbitals on the same Ni site, we further evaluate
\begin{equation}
C_{xz}=\frac{1}{N_s}\sum_{i,\ell}
\left\langle m^z_{i\ell x}m^z_{i\ell z}\right\rangle,
\qquad
m^z_{i\ell\alpha}=n_{i\ell\alpha\uparrow}-n_{i\ell\alpha\downarrow}.
\label{eq:Cxz}
\end{equation}
A positive $C_{xz}$ corresponds to a tendency toward parallel spin alignment between the $x$ and $z$ orbitals.

To provide a noninteracting reference for the effects of the underlying band structure, we calculate the bare static Lindhard susceptibility~\cite{Graser2009,Xia2025},
\begin{equation}
	\begin{aligned}
		\chi_0^P(\bm q)
		={}&-\frac{1}{N_k}\sum_{\bm k,m,n}
		\frac{f(\xi_{m\bm k})-f(\xi_{n,\bm k+\bm q})}
		{\xi_{m\bm k}-\xi_{n,\bm k+\bm q}}
		\\
		&\times
		\left|\langle u_{m\bm k}|P|
		u_{n,\bm k+\bm q}\rangle\right|^2,
	\end{aligned}
	\label{eq:chi0}
\end{equation}
where $\xi_{m\bm k}=E_{m\bm k}-\mu$, and $E_{m\bm k}$ and $|u_{m\bm k}\rangle$ are the eigenvalues and eigenvectors of the noninteracting Bloch Hamiltonian, respectively. All interaction terms are set to zero, while the hopping parameters, crystal-field splitting, and interlayer potential bias are retained. We use the discrete momenta allowed by periodic boundary conditions. Results for $L=8$ and $L=20$ are compared at $\Delta E/t=0.8$ and $k_BT/t=0.005$. The Fermi function is $f(\xi)=[\exp(\xi/k_BT)+1]^{-1}$. The chemical potential is adjusted at each interlayer bias to maintain a spin-summed average filling of $\langle n\rangle=0.75$ over all spatial orbitals.

In the basis $(1x,1z,2x,2z)$, we use $P_{\mathrm{tot}}=I_4$, $P_x=\mathrm{diag}(1,0,1,0)$, $P_z=\mathrm{diag}(0,1,0,1)$, and $P_{2x}=\mathrm{diag}(0,0,1,0)$ to obtain the total response, the responses projected onto the $x$ and $z$ orbital sectors, and the lower-layer $x$-orbital response, respectively. The susceptibility is evaluated at $\bm M=(\pi,\pi)$ and reported per spin and per bilayer unit cell.

The ground-state properties are calculated using the CPMC method~\cite{Zhang1995,Zhang1997}. In this approach, the ground-state wave function is projected from a trial wave function $|\Psi_T\rangle$, which has a nonzero overlap with the ground state, by an importance-sampled branching random walk in an overcomplete space of Slater determinants,
\begin{equation}
	|\Psi_0\rangle \propto
	\lim_{\Theta\rightarrow\infty} e^{-\Theta H}|\Psi_T\rangle .
\end{equation}
After a Hubbard--Stratonovich transformation of the interaction terms, the projection is realized by an ensemble of walkers $|\phi\rangle$, each being a Slater determinant, undergoing importance-sampled random walks in imaginary time.

For doped multiorbital Hubbard models, auxiliary-field quantum Monte Carlo suffers from the fermion sign problem~\cite{Troyer2005}. In CPMC, this problem is controlled by the constrained-path approximation,
\begin{equation}
	\langle\Psi_T|\phi\rangle>0 ,
\end{equation}
for each walker $|\phi\rangle$, so that every walker retains a positive overlap with the trial wave function throughout the random walk.

Extensive benchmark studies have shown that this method provides accurate ground-state energies and correlation functions in Hubbard models: the systematic error induced by the constraint is typically within a few percent, and the ground-state observables are largely insensitive to the choice of the trial wave function~\cite{Zhang1997}. Nevertheless, the constraint introduces a systematic dependence on the trial wave function~\cite{Qin2016,Xu2022}. We therefore focus mainly on robust trends across parameter scans. In the present calculations, the trial wave function is chosen as the ground state of the corresponding noninteracting Hamiltonian at the same filling. We use an imaginary-time step $\Delta\tau=0.02t^{-1}$ and parallelize the simulations using the Message Passing Interface, with 16 processes and 1200 walkers per process, giving a total population of 19200 walkers. The resulting statistical error bars are smaller than the symbol size unless visible in the figures. The equal-time observables are evaluated using back propagation to reduce mixed-estimator bias.

\section{Results and Discussion}

\begin{figure}[!htbp]
	\vspace{0.2cm}
	\centering
	\includegraphics[width=0.48\textwidth]{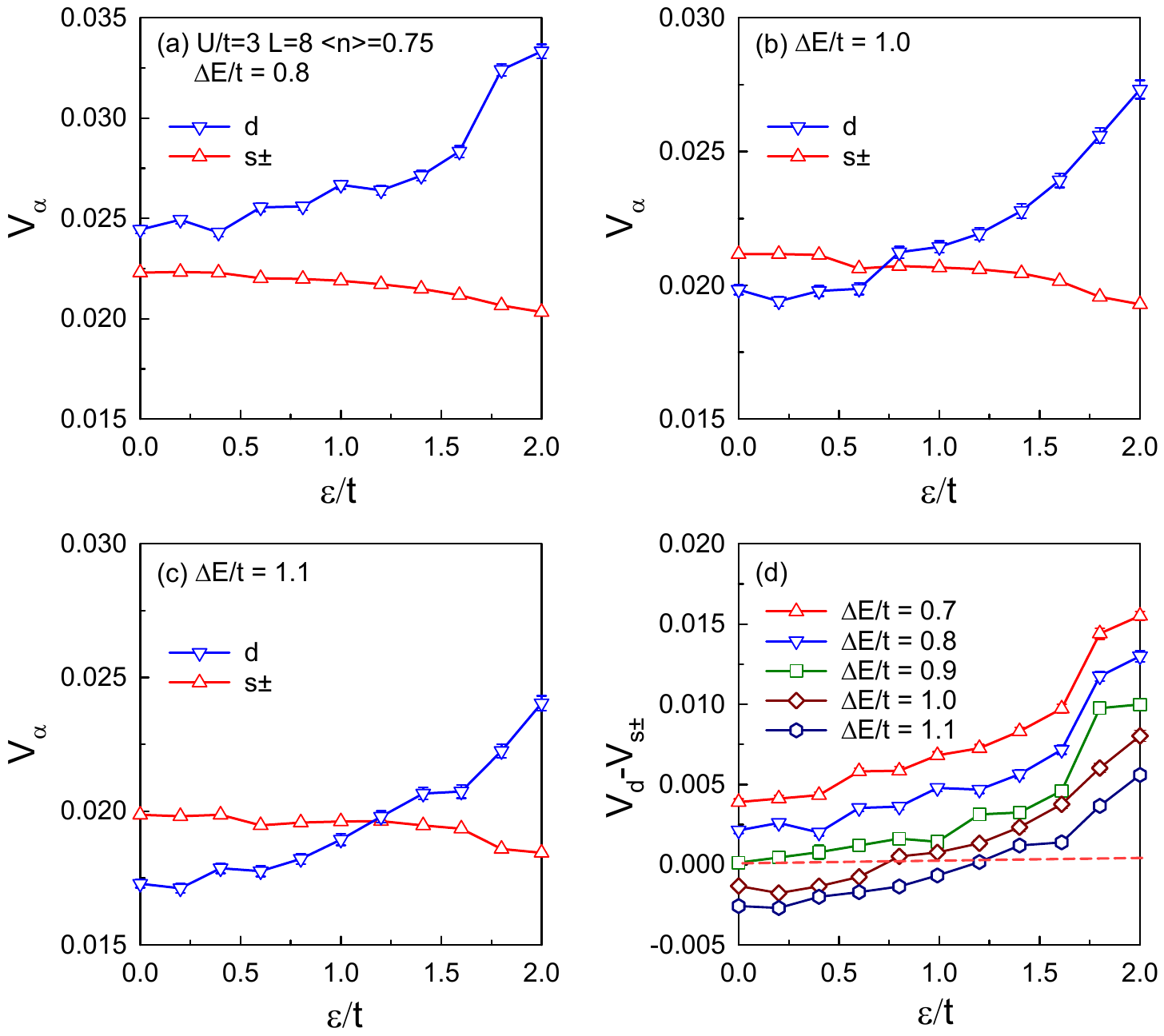}
	\caption{Evolution of the pairing vertex contributions with the electric-field potential difference $\varepsilon$ at $U/t=3$, $J_H=0$, $L=8$, and $\langle n\rangle=0.75$. (a)--(c) $V_d$ and $V_{s^{\pm}}$ for representative crystal-field splittings. (d) $\Delta V=V_d-V_{s^{\pm}}$ for $\Delta E/t=0.7$--1.1. Increasing $\varepsilon$ systematically favors the $d$-wave channel relative to $s^{\pm}$, whereas increasing $\Delta E$ shifts the leading-pairing crossover to larger $\varepsilon$.}
	\label{fig:field_deltaE}
	\vspace{0.2cm}
\end{figure}

Figure~\ref{fig:field_deltaE} shows the electric-field potential difference response at $U/t=3$ and $J_H=0$. For all crystal-field splittings shown, increasing $\varepsilon$ enhances $V_d$ while gradually suppressing $V_{s^{\pm}}$. At $\Delta E/t=0.8$, the $d$-wave vertex is already larger at weak electric-field potential difference and its advantage increases overall with $\varepsilon$. At larger $\Delta E/t=1.0$ and 1.1, the $s^{\pm}$ channel becomes more competitive near zero field, but the opposite field dependences of the two vertex contributions eventually drive a crossover toward $d$-wave dominance. The difference $\Delta V=V_d-V_{s^{\pm}}$ shown in Fig.~\ref{fig:field_deltaE}(d) directly reflects the competition between the two pairing channels. $\Delta V$ increases with $\varepsilon$, while increasing $\Delta E$ lowers $\Delta V$ and shifts the crossover to a larger layer potential difference. In the range shown, a larger $\Delta E$ suppresses both pairing vertex contributions, with a stronger suppression in the $d$-wave channel.

\begin{figure}[!htbp]
	\vspace{0.2cm}
	\centering
	\includegraphics[width=0.48\textwidth]{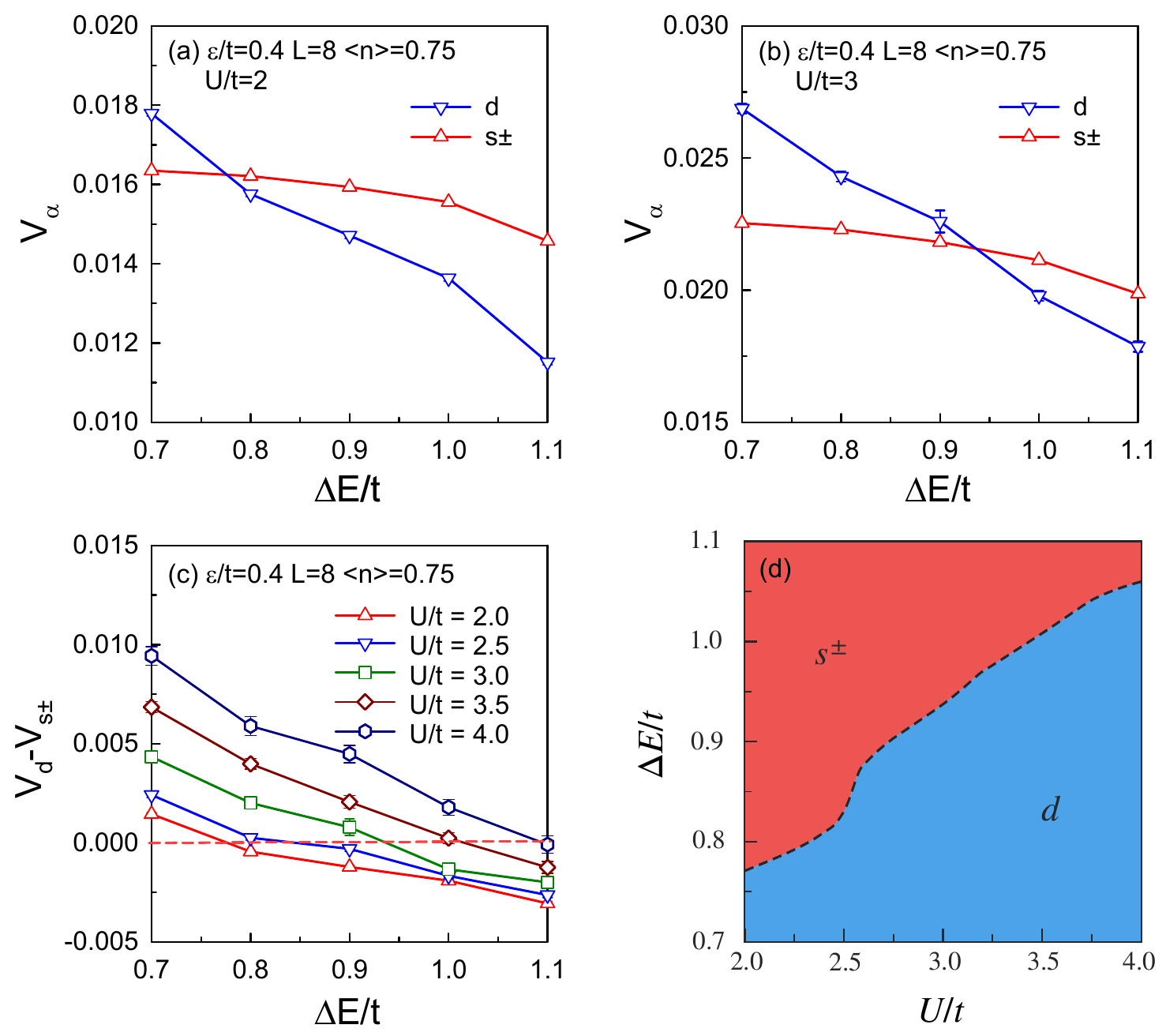}
	\caption{Pairing competition as a function of Hubbard interaction and crystal-field splitting at $\varepsilon/t=0.4$, $J_H=0$, $L=8$, and $\langle n\rangle=0.75$. (a) and (b) Pairing vertex contributions $V_d$ and $V_{s^{\pm}}$ versus $\Delta E/t$ for $U/t=2$ and 3, respectively. (c) $\Delta V=V_d-V_{s^{\pm}}$ for different $U$. The horizontal dashed line marks $\Delta V=0$. (d) Dominant pairing tendencies in the $(U,\Delta E)$ plane. Increasing $U$ favors the $d$-wave channel relative to the $s^{\pm}$ channel, whereas increasing $\Delta E$ makes the $s^{\pm}$ channel relatively more competitive.
	}
	\label{fig:U}
	\vspace{0.2cm}
\end{figure}

The role of the onsite Hubbard interaction is shown in Fig.~\ref{fig:U}. At fixed $\varepsilon/t=0.4$, increasing $U/t$ from 2 to 3 enhances both pairing vertex contributions, with a larger increase in $V_d$, as shown in Figs.~\ref{fig:U}(a) and (b). Figure~\ref{fig:U}(c) further shows that $\Delta V=V_d-V_{s^{\pm}}$ shifts upward as $U/t$ increases from 2 to 4, whereas increasing $\Delta E$ makes the $s^{\pm}$ channel relatively more competitive. As shown in Fig.~\ref{fig:U}(d), the boundary defined by $V_d=V_{s^{\pm}}$ moves toward larger $\Delta E/t$ as $U$ increases, expanding the region in which the $d$-wave vertex is larger. Within this parameter regime, $U$ preferentially strengthens the intralayer $d$-wave tendency and partly compensates the crystal-field-induced suppression of the $x$-orbital channel.

\begin{figure}[!htbp]
\vspace{0.2cm}
\centering
\includegraphics[width=0.48\textwidth]{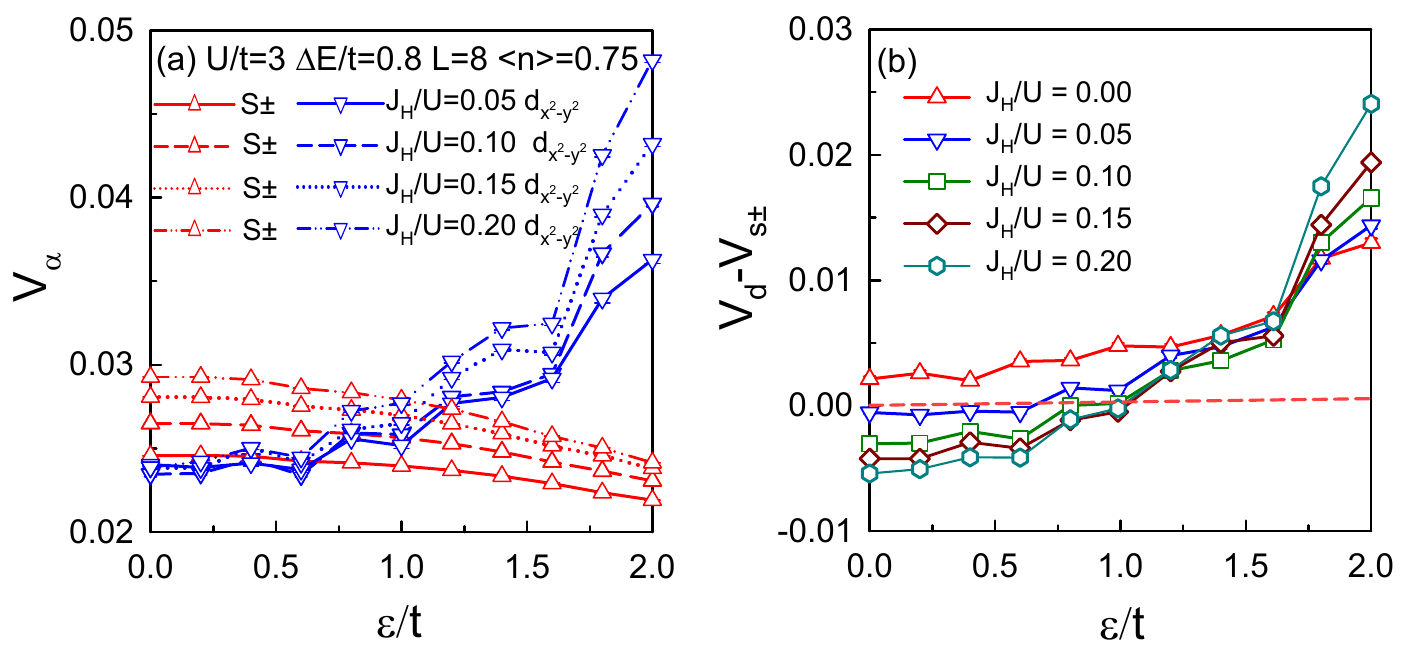}
\caption{Dependence of pairing competition on $J_H$ at $U/t=3$, $\Delta E/t=0.8$, $L=8$, and $\langle n\rangle=0.75$. (a) Pairing vertex contributions $V_d$ and $V_{s^{\pm}}$ versus $\varepsilon/t$ for $J_H/U=0.05$, 0.10, 0.15, and 0.20. (b) $\Delta V=V_d-V_{s^{\pm}}$ for several $J_H/U$. The horizontal dashed line marks $\Delta V=0$. Increasing $J_H$ makes the $s^{\pm}$ channel relatively more competitive at weak $\varepsilon$, while increasing the relative strength of the $d$-wave channel at large $\varepsilon$.}
\label{fig:Hund}
\vspace{0.2cm}
\end{figure}

The preceding results were obtained at $J_H=0$. We next examine the dependence on $J_H$ at fixed $U/t=3$ and $\Delta E/t=0.8$, as shown in Fig.~\ref{fig:Hund}. Fig.~\ref{fig:Hund}(a) shows that the pairing vertex contributions remain dependent on $J_H$ across the $\varepsilon$ range. At weak layer potential difference, increasing $J_H$ makes the $s^{\pm}$ channel relatively more competitive, as reflected by the more negative $\Delta V=V_d-V_{s^{\pm}}$ in Fig.~\ref{fig:Hund}(b). At large layer potential difference, however, $V_d$ increases more strongly with $J_H$, and $\Delta V$ becomes more positive. For the finite $J_H$ values shown, increasing $\varepsilon$ changes the relative dominance from $s^{\pm}$ to $d$-wave.

The phase diagram in Fig.~\ref{fig:model}(a) shows that in the region of small electric-field potential difference, increasing $J_H$ initially extends the $s^{\pm}$-dominant region toward larger $\varepsilon$. In the region of large electric-field potential difference, however, the $d$-wave vertex contribution remains larger throughout the investigated $J_H$ range, indicating that increasing Hund's coupling is insufficient to restore the relative dominance of the $s^{\pm}$ channel in this regime. A possible interpretation involves the interplay between Hund-induced magnetic correlations and layer imbalance. Hund's coupling favors parallel spin alignment between the two orbitals on the same Ni site, thereby linking their magnetic responses. Previous work has shown that this coupling can transmit the interlayer antiferromagnetic correlations of the $d_{3z^2-r^2}$ orbitals to the $d_{x^2-y^2}$ orbitals, promoting interlayer spin-singlet pairing in the latter~\cite{QuHund2025}. Although this mechanism concerns a different orbital pairing channel, it illustrates how Hund's coupling can promote interlayer pairing through coupled orbital magnetic correlations and provides a possible context for the expansion of the $s^{\pm}$-dominant region observed here.

As $\varepsilon$ increases, charge redistribution makes the two layers increasingly inequivalent. The resulting layer imbalance may reduce the relative advantage of interlayer pairing over intralayer pairing, consistent with studies of electric-field tuning in bilayer nickelates~\cite{Shao2026}. This competition offers a possible explanation for the limited expansion of the $s^{\pm}$-dominant region at larger $\varepsilon$. The nearly vertical boundary indicates that the crossing point between the two pairing channels becomes weakly dependent on $J_H$, rather than demonstrating saturation of the individual pairing vertex contributions. Establishing the microscopic origin of this behavior requires evidence beyond the pairing vertex contribution alone.

\begin{figure}[!htbp]
\centering
\includegraphics[width=0.48\textwidth]{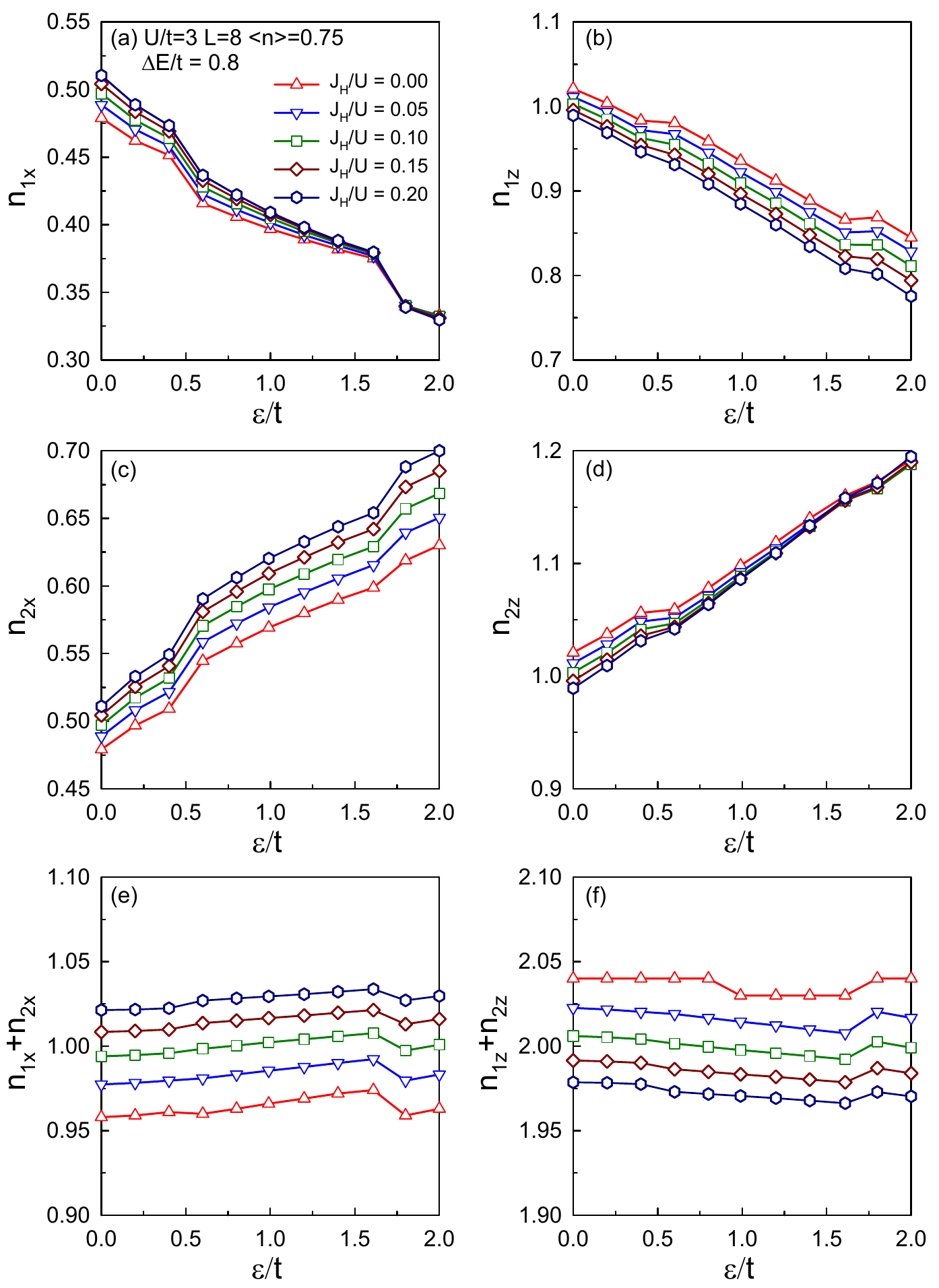}
\caption{Orbital occupations versus $\varepsilon$ for different $J_H$ at $U/t=3$, $\Delta E/t=0.8$, $L=8$, and $\langle n\rangle=0.75$. Panels (a)--(d) show $n_{1x}$, $n_{1z}$, $n_{2x}$, and $n_{2z}$, respectively, while panels (e) and (f) show $n_{1x}+n_{2x}$ and $n_{1z}+n_{2z}$. Here, $n_{\ell\alpha}$ denotes the site-averaged, spin-summed occupation of orbital $\alpha=x,z$ in layer $\ell=1,2$. Increasing $\varepsilon$ redistributes charge from layer 1 to layer 2, while the total occupation of each orbital remains nearly unchanged. Increasing $J_H$ shifts charge from the $z$ sector toward the $x$ sector, reducing orbital polarization.}
\label{fig:density}
\end{figure}

The microscopic origin of these trends can be further examined from the orbital occupations shown in Fig.~\ref{fig:density}. A positive $\varepsilon$ raises the onsite energy of layer 1 and lowers that of layer 2. Accordingly, $n_{1x}$ and $n_{1z}$ decrease, while $n_{2x}$ and $n_{2z}$ increase. Meanwhile, the sums $n_{1x}+n_{2x}$ and $n_{1z}+n_{2z}$ remain nearly constant as $\varepsilon$ varies. Thus, in the moderate-correlation regime studied here, the field response is primarily characterized by same-orbital charge transfer between the two layers, with no substantial net redistribution between the $x$ and $z$ orbital sectors. 

Fig.~\ref{fig:density} also shows that the orbital occupations depend on $J_H$. At fixed $\varepsilon$, increasing $J_H$ raises the total $x$-orbital occupation and reduces the total $z$-orbital occupation, thereby weakening the orbital polarization associated with the crystal-field splitting. The same-spin interorbital interaction is $U-3J_H$, so larger $J_H$ lowers the energetic cost of distributing electrons over the two orbitals with parallel spins. Thus, the occupation data reveal that increasing $\varepsilon$ mainly redistributes charge between layers, whereas increasing $J_H$ changes its distribution between orbitals. Recent dynamical cluster approximation quantum Monte Carlo studies have shown that layer-density imbalance itself can substantially modify the competition between $d$-wave and $s^{\pm}$ pairing in bilayer Hubbard models, even without incorporating the specific orbital structure of bilayer nickelates~\cite{Jia2026}. This suggests that layer imbalance can serve as an independent tuning parameter for the competition between different pairing channels.

\begin{figure}[!htbp]
\centering
\includegraphics[width=0.48\textwidth]{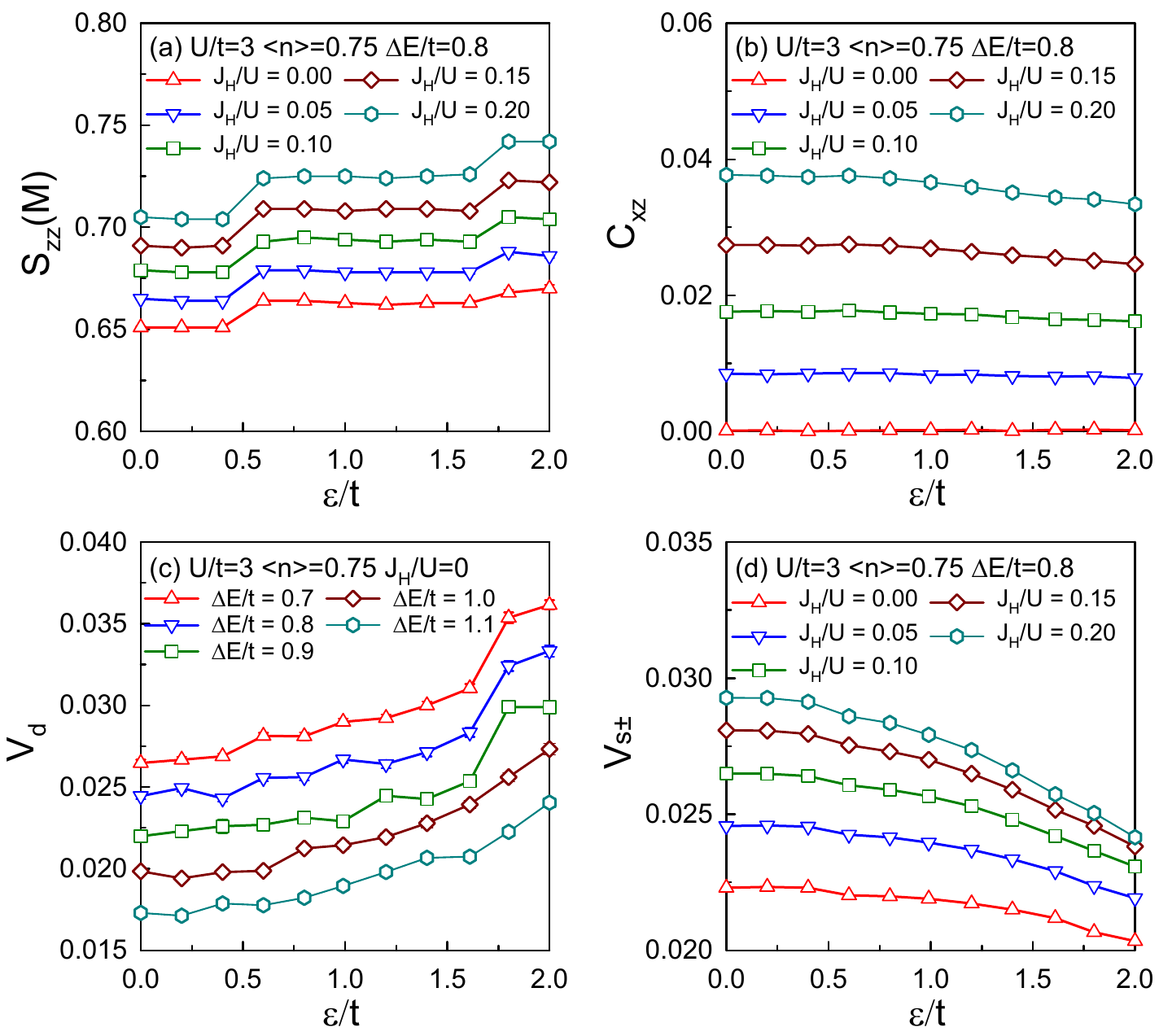}
\caption{Magnetic and pairing vertex contributions at $U/t=3$, $L=8$, and $\langle n\rangle=0.75$. (a) Antiferromagnetic spin structure factor $S^{zz}(M)$ for several $J_H/U$ at $\Delta E/t=0.8$. (b) Onsite interorbital Hund spin correlation $C_{xz}$ defined in Eq.~\eqref{eq:Cxz}. (c) $d$-wave pairing vertex contribution for several crystal-field splittings. (d) Interlayer $s^{\pm}$ pairing vertex contribution for several values of $J_H/U$. The electric-field potential difference enhances the antiferromagnetic and $d$-wave responses, while Hund's coupling increases the local $x$--$z$ parallel-spin correlation.}
\label{fig:magnetic}
\end{figure}

As shown in Fig.~\ref{fig:density}(c), $n_{2x}$ moves progressively toward the single-orbital half-filled value $n_x=1$ with increasing $\varepsilon$. In a Hubbard-like square-lattice sector this favors stronger $(\pi,\pi)$ antiferromagnetic correlations. Magnetic and correlation studies of La$_3$Ni$_2$O$_7$ also emphasize sizable interlayer exchange and pronounced orbital dependence of the spin response~\cite{ChenRIXS2024,Botzel2024,LechermannPRM2024}. Consistently, Fig.~\ref{fig:magnetic}(a) shows that the in-plane antiferromagnetic spin structure factor $S^{zz}(M)$ increases with electric-field potential difference, while Fig.~\ref{fig:magnetic}(c) shows a simultaneous enhancement of $V_d$. These results suggest a possible connection between the increase of $n_{2x}$, the enhancement of intralayer antiferromagnetic correlations, and the strengthened $d$-wave pairing tendency under an electric-field potential difference. The same figure also shows that the in-plane antiferromagnetic spin structure factor $S^{zz}(M)$ increases with $J_H$. This reflects both the $J_H$-induced increase of the $x$-orbital occupation and the tendency of Hund's coupling to stabilize larger local moments.

\begin{figure}[!htbp]
\centering
\includegraphics[width=0.45\textwidth]{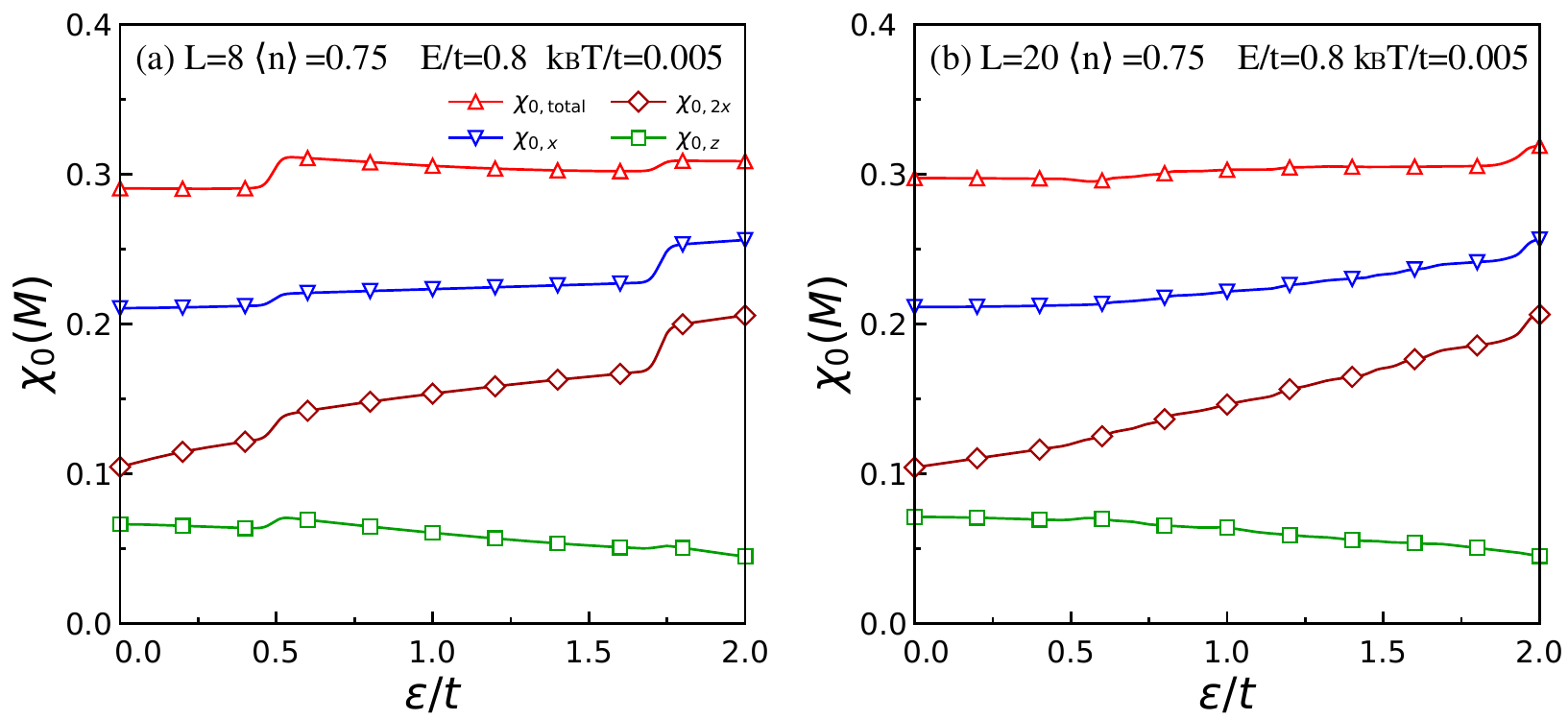}
\caption{Bare static susceptibility at $M=(\pi,\pi)$ as a function of the interlayer potential bias $\varepsilon$ for (a) $L=8$ and (b) $L=20$, with $\langle n\rangle=0.75$, $\Delta E/t=0.8$, and $k_BT/t=0.005$. The curves show the total response $\chi_{0,\mathrm{total}}$, the responses projected onto the $x$ and $z$ orbitals of both layers, $\chi_{0,x}$ and $\chi_{0,z}$, and the response projected onto the layer-2 $x$ orbital, $\chi_{0,2x}$ . The $L=8$ results exhibit step-like variations. These features become less pronounced upon increasing the system size to $L=20$.}
\label{fig:chi0}
\end{figure}

Meanwhile, Fig.~\ref{fig:magnetic}(a) shows that the $S^{zz}(M)$ curves exhibit step-like enhancements near the field intervals $\varepsilon/t\simeq0.4$--0.6 and 1.6--1.8. Similar variations in the lower-layer $x$-orbital occupation $n_{2x}$ occur in approximately the same field intervals, as shown in Fig.~\ref{fig:density}(c), suggesting a possible connection between charge redistribution and the magnetic response. To determine whether these features may be related to the underlying noninteracting electronic structure, we calculate the bare static susceptibility defined in Eq.~\eqref{eq:chi0} for $L=8$ and $L=20$ at $k_BT/t=0.005$. Fig.~\ref{fig:chi0}(a) shows that the $L=8$ results exhibit step-like variations in similar intervals, particularly in the total and lower-layer $x$-orbital responses. These features are substantially reduced for $L=20$, as shown in Fig.~\ref{fig:chi0}(b). These observations suggest that bias-induced changes in electronic occupations may influence the magnetic response and contribute to the similar features in $n_{2x}$, $\chi_0(M)$, and $S^{zz}(M)$. The size dependence of $\chi_0(M)$ further suggests that finite-size electronic structure may play a role in this behavior.

The Hund-coupling dependence is further illustrated in Fig.~\ref{fig:magnetic}(b). At fixed $\varepsilon$, $C_{xz}$ increases nearly monotonically with $J_H$, consistent with enhanced local parallel-spin alignment between the $\dx$ and $\dz$ orbitals. $V_{s^{\pm}}$ shows a similar tendency, as seen in Fig.~\ref{fig:magnetic}(d). Previous theoretical studies have discussed how Hund's coupling can connect the magnetic responses of the two orbitals and promote interlayer spin-singlet pairing~\cite{Oh2023,Lu2024,Tian2024,Xiong2026}. These studies provide a possible context for the increased relative competitiveness of the $s^{\pm}$ channel at weak electric-field potential difference. However, the local correlation $C_{xz}$ alone does not establish the transfer of interlayer magnetic exchange to the pairing channel in the present model.

The occupation data reveal how the electronic environment changes as the electric-field potential difference increases. The growing difference $n_{2z}-n_{1z}$ indicates increasing layer imbalance in the $z$ sector, which may weaken interlayer pairing relative to intralayer pairing. This interpretation is qualitatively consistent with previous studies of pairing in layer-imbalanced bilayer systems~\cite{Jia2026,Wei2026}. Meanwhile, the $x$-orbital occupation in the lower-potential layer approaches half filling, strengthening the intralayer antiferromagnetic response and $V_d$. 

Taken together, the results show that the pairing competition responds differently to layer polarization, orbital polarization, and local interorbital spin correlations. Increasing $U$ enhances both pairing vertex contributions, with a larger enhancement in the $d$-wave channel, whereas increasing $\Delta E$ makes the $s^{\pm}$ channel relatively more competitive. Hund's coupling increases the relative advantage of $s^{\pm}$ pairing at weak electric-field potential difference, while $d$-wave pairing dominates at sufficiently large electric-field potential difference. 

\section{Conclusion}

We have used ground-state CPMC calculations to examine the competition between interlayer $s^{\pm}$ and intralayer $d$ pairing tendencies in an electrically imbalanced bilayer two-orbital Hubbard model for La$_3$Ni$_2$O$_7$. Within the parameter range studied, increasing the effective layer potential difference reduces $V_{s^{\pm}}$ and increases $V_d$. Increasing the crystal-field splitting reduces both vertex contributions, with a larger effect on the $d$-wave channel, whereas increasing $U$ enhances both interaction-induced pairing contributions and has a stronger effect on $V_d$. Hund's coupling mainly increases the relative competitiveness of $s^{\pm}$ pairing in the weak-field regime and changes the crossover field only moderately over the range considered.

The charge and magnetic observables provide a qualitative picture of these trends. The electric-field potential difference mainly redistributes charge between the two layers within each orbital sector and moves the lower-potential-layer $\dx$ orbital toward half filling. This evolution is accompanied by increases in the in-plane antiferromagnetic static spin structure factor  $S^{zz}(M)$ and the $d$-wave pairing vertex contribution.  The interacting $S^{zz}(M)$ and the bare susceptibility for $L=8$ exhibit similar step-like variations in approximately the same field intervals. These features become less pronounced in the bare susceptibility for $L=20$, suggesting that finite-size electronic structure may contribute to the observed magnetic response. Hund's coupling increases the onsite interorbital spin correlation $C_{xz}$, consistent with stronger local parallel-spin alignment. In the presence of substantial $\dz$ interlayer hopping, these observations suggest a possible role for coupled orbital magnetic correlations in the relative competitiveness of interlayer $s^{\pm}$ pairing at weak layer polarization. At larger bias, increasing layer imbalance may reduce the relative advantage of interlayer pairing, consistent with the dominance of the $d$-wave channel in the present finite-size calculations.

Overall, the calculations indicate that the relative $s^{\pm}$ and $d$-wave pairing tendencies depend on the combined effects of layer polarization, crystal-field-induced orbital polarization, Hubbard interaction, and Hund's coupling.  These conclusions are based on the selected pairing vertex contributions in finite systems and are subject to the trial-state-dependent constrained-path approximation. These results provide insight into how electric-field-induced layer imbalance and multiorbital interactions can modify the competition between different pairing  channels in bilayer nickelate models. The electric-field-induced layer polarization studied here offers a complementary means of manipulating orbital occupations and electronic coupling without changing the chemical composition or introducing extrinsic disorder. The connection between layer redistribution, magnetic correlations, and pairing competition may inform future experiments aimed at identifying the microscopic origin of superconductivity in multiorbital nickelates.

\begin{acknowledgments}
This work was supported by the National Natural Science Foundation of China under Grants No. 12474218 and the Beijing Natural Science Foundation under Grants No. 1242022 and No. 1252022. Numerical simulations were carried out at the High-Performance Scientific Computing Center of Beijing Normal University.
\end{acknowledgments}

\section*{Data Availability}
The data that support the findings of this study are available from the corresponding author upon reasonable request.

\bibliography{references}

\end{document}